\documentclass[twocolumn,10pt,oneside,reqno]{article}
\usepackage[letterpaper,top=1in, bottom=1in, left=.65in, right=.65in]{geometry}
\usepackage{amsmath, amsthm, amssymb,mathrsfs}
\usepackage[normalem]{ulem}
\usepackage{graphicx}
\usepackage{epstopdf}
\usepackage{hyperref}
\usepackage{overpic}
\usepackage{cancel}
\usepackage{rotating}
\usepackage{url}
\usepackage{caption}
\usepackage{color}
\usepackage[usenames,dvipsnames,svgnames,table]{xcolor}
\usepackage{booktabs}
\usepackage{rotating}
\usepackage{multirow}
\usepackage{wrapfig}
\usepackage{multicol}
\usepackage{multirow}
\usepackage{adjustbox}
\usepackage{setspace}
\usepackage{palatino} 
\usepackage{algorithm,algcompatible}
\usepackage{algpseudocode}
\usepackage{tikz}
\usepackage{booktabs}
\usepackage{float}
\usepackage{algorithm}
\usepackage{algpseudocode}
\renewcommand{\Comment}[2][.4\linewidth]{%
	\leavevmode\hfill\makebox[#1][l]{$\triangleright$~#2}}
\usepackage{mathtools}  
\usepackage{psfrag}
\usepackage{tikz}
\usepackage[T1]{fontenc}
\usepackage{array}
\usepackage{makecell}
\usepackage{afterpage}
\newcolumntype{x}[1]{>{\centering\arraybackslash}p{#1}}

\usepackage[bottom,flushmargin,hang,multiple]{footmisc}
\usepackage{lipsum}

\makeatletter
\renewcommand\footnoterule{%
  \kern-3\p@
  \hrule\@width5.5cm
  \kern2.6\p@}
\makeatother
  
\algnewcommand\algorithmicinput{\textbf{Input:}}
\algnewcommand\INPUT{\item[\algorithmicinput]}
\algnewcommand\algorithmicoutput{\textbf{Output:}}
\algnewcommand\OUTPUT{\item[\algorithmicoutput]}
\algnewcommand\algorithmicoptional{\textbf{Optional:}}
\algnewcommand\OPTIONAL{\item[\algorithmicoptional]}

\graphicspath{{Figures/}}
\newcommand{\boundellipse}[3]
{(#1) ellipse (#2 and #3)
}
\newcommand{\overbar}[1]{\mkern 6mu\overline{\mkern-6mu#1\mkern-6mu}\mkern 6mu}

\title{\huge{FTL: Transfer Learning Nonlinear Plasma Dynamic Transitions in Low Dimensional Embeddings via Deep Neural Networks}
}

\author{Zhe Bai$^{1,*}$, Xishuo Wei$^{2}$, William Tang$^{3,4}$, Leonid Oliker$^{1}$, Zhihong Lin$^{2}$, Samuel Williams$^{1}$\\~\\
\small{$^1$ Applied Mathematics and Computational Research Division, Lawrence Berkeley National Lab, Berkeley, CA 98195, United States}\\
\small{$^2$ Department of Physics and Astronomy, University of California, Irvine, CA 92697, United States}\\
\small{$^3$ Princeton Plasma Physics Laboratory, Princeton, NJ 08543, United States}\\
\small{$^4$Princeton University, Princeton, NJ 08544, United States}
\\~\\
}
\date{}

\begin{document}

\twocolumn[
  \begin{@twocolumnfalse}

\maketitle
\begin{abstract}
Deep learning algorithms provide a new paradigm to study high-dimensional dynamical behaviors, such as those in fusion plasma systems.  Development of novel model reduction methods, coupled with detection of abnormal modes with plasma physics, opens a unique opportunity for building efficient models to identify plasma instabilities for real-time control. Our Fusion Transfer Learning (FTL) model demonstrates success in reconstructing nonlinear kink mode structures by learning from a limited amount of nonlinear simulation data. The knowledge transfer process leverages a pre-trained neural encoder-decoder network, initially trained on linear simulations, to effectively capture nonlinear dynamics. The low-dimensional embeddings extract the coherent structures of interest, while preserving the inherent dynamics of the complex system. Experimental results highlight FTL's capacity to capture transitional behaviors and dynamical features in plasma dynamics --- a task often challenging for conventional methods. The model developed in this study is generalizable and can be extended broadly through transfer learning to address various magnetohydrodynamics (MHD) modes.\\

\end{abstract}
~\\
  \end{@twocolumnfalse}]
  
\let\thefootnote\relax
\footnotetext{
\hspace{10pt}$^*$ Corresponding author (zhebai@lbl.gov)}
\fontsize{10}{12}\selectfont


\section{Introduction}
Modeling, design, and control of fusion plasma that characterize a high-dimensional, strong nonlinear system are a central challenge for tokamak-based plasma devices. In such systems, the wide range of scales in both space and time necessitates a large amount of data collection, processing, and computation to resolve all the physical relevant features for system identification and control. The intricate, multi-scale dynamics usually render classic control methods impractical for real-time feedback control of tokamak experiments. Dynamic model-based approaches provide a viable alternative, enabling the implementation of more adaptive and robust control strategies. However, their efficacy relies on the availability of mathematical models to accurately emulate the system dynamics. Consequently, computational modeling plays an essential role in understanding, estimating and eventually developing model-based control for such complex physics processes.

Characterizing the evolution of plasma dynamics is crucial for effective plasma control. In magnetically-confined plasmas, the macroscopic magnetohydrodynamic (MHD) instabilities can be excited by equilibrium currents or pressure gradients. The various MHD instabilities, including fishbones, sawteeth, neoclassical tearing modes (NTM), and kink modes, can limit burning plasma performance and threaten fusion device integrity\cite{McGuire1983Fishbone,Goeler1974Sawtooth,Blank2008MHD,Carrera1986NTM}. 
Although the kink mode does not directly cause confinement degradation, it can non-linearly excite the more destructive modes like sawtooth and NTM that affect the particle confinement or even cause a major plasma disruption.
Detecting anomalies allows for early identification of the critical events, enabling proactive measures to mitigate them and maintain stable plasma conditions.
Investigations of plasma instabilities can enhance understanding and prediction of core plasma dynamics and transport, as well as edge plasma-material interactions critical for operation of future fusion devices like ITER~\cite{ITER_Physics_Chap1,ITER_physics_Chap2}.
The evolution of instabilities typically encompasses of linear and nonlinear phases, with the former characterized by small perturbations and the later with saturated amplitude.
As perturbation amplitude increases exponentially in the linear phase, nonlinear damping effects compete against the linear instability drive, leading to the decrease of the growth rate and eventually the saturation of the perturbation amplitude, with fluctuation amplitudes reaching a steady level, quenching to near-zero, or exhibiting limit cycle oscillation patterns~\cite{Chen-PoP2000,diamond2005zonal}.
In fusion experiments, directly observing the linear growing phase is challenging. Understanding the nonlinear evolution of fluctuations induced by instabilities is crucial for validating simulations, predicting physical dynamics, and preventing plasma disruptions.
Nonlinear bifurcations in the dynamics of the plasma are frequently observed, exemplified by transitions such as from low-confinement mode (L-mode) to high-confinement mode (H-mode) induced by auxiliary heating, variations in turbulence-zonal flow interactions due to distinct damping parameters, and the formation of internal transport barriers resulting from the saturation of macroscopic MHD modes.
The bifurcations signify the existence of multiple steady states of plasma turbulence and often indicate pathways to enhanced confinement. Such bifurcations appear at the nonlinear saturation of unstable waves, driven by reductions in free energy and increased damping effects.


Plasma instabilities such as MHD modes, micro-instabilities, and Alfven eigenmodes are extensively studied using physics models and computational algorithms, 
including MHD codes~\cite{JARDIN2004M3DC1,Sovinec_2004,king2016nimrod,ferraro2006finite,charlton1986numerical,charlton1990compressible}, kinetic particle-in-cell codes~\cite{lin1998turbulent, chang2004numerical,ku2009full,chen2007electromagnetic}, and kinetic Eulerian codes~\cite{candy2003eulerian, jenko2000electron,goerler2011global}, among others.
Gyrokinetic Toroidal Code (GTC)~\cite{lin1998turbulent} stands as a leading plasma physics simulation tool, incorporating multiple simulation models such as gyrokinetic, fully-kinetic, and fluid covering multi-scale physics. The validation of multi-scale nonlinear simulations can be quantitatively achieved through their comparison with experimental data~\cite{Liu2022PRL, Brochard2024PRL}.


\begin{figure*}[htb!]
\begin{center}
\includegraphics[width=.99\textwidth]{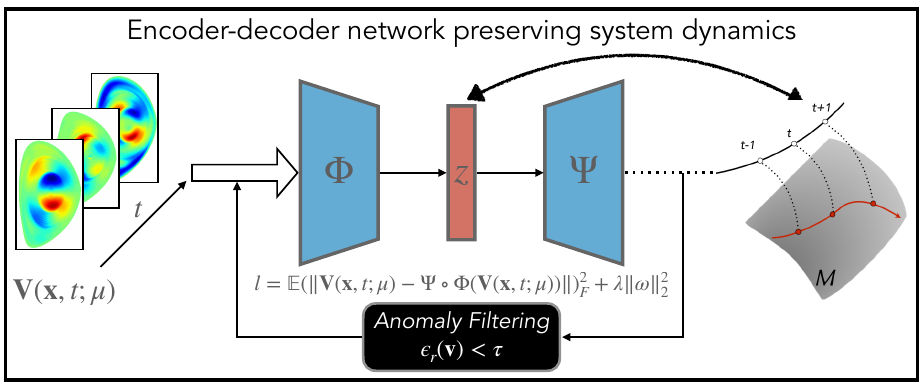}
\caption{Schematic diagram of the FTL model using an encoder ($\Phi$) - decoder ($\Psi$) network. The coherent structure is compressed to a low-dimensional vector $\boldsymbol{Z}$ while the inherent dynamics are preserved in the latent space through embedding. Regularization is employed to impose penalties on the model's complexity. An anomaly filtering subroutine is utilized for the identification of outliers, thereby improving the fidelity of the pre-trained model.} \label{fig01}
\end{center}
\end{figure*}

High-fidelity simulations of systems governed by nonlinear partial differential equations (PDEs) often entail substantial computational costs, making them impractical for decision-making tasks in real-world applications. To address this issue, model order reduction methods have been developed to approximate the behaviors of such system while significantly reducing computational overhead. 
Reduced-order models (ROMs) play a pivotal role in characterizing, estimating, and controlling high-dimensional complex systems~\cite{lucia2004reduced,quarteroni2014reduced,keiper2018reduced,choi2020gradient}. Although the fully-resolved state space of a complex system, e.g., turbulence and tokamak plasmas, may contain millions of degrees of freedom, these nonlinear systems exhibit inherent low-dimensional, dominant patterns in latent space, and typically evolve on a low-dimensional attractor that can be leveraged for effective control~\cite{noack2011reduced, duriez2017machine}. The essential steps in model reduction include identifying an optimal state space for the attractor and characterizing the system's dynamical behavior when evolved on this attractor. Conventional approaches are projection-based, such as POD-Galerkin method that performs Galerkin projection of first principle equations onto a linear subspace of modes obtained via proper orthogonal decomposition (POD)~\cite{holmes2012turbulence, benner2015survey, taira2017modal}. 
Dynamic mode decomposition (DMD) is another dimensionality reduction technique to decompose high-dimensional data into dominant spatiotemporal coherent structures~\cite{Schmid2010jfm,Rowley2009jfm}. As a numerical technique to approximate the Koopman operator~\cite{Koopman1931pnas,Mezic2005nd,Bagheri2013jfm}, DMD and its variations~\cite{proctor2016dynamic,bai2020dynamic} have established a strong connection to the analysis of nonlinear dynamical systems by constructing modes that oscillate at a fixed frequency and growth or decay rate in a linear evolution model.
Non-intrusive model reduction based on projection methods enables approximating the low-dimensional operators using regression techniques~\cite{benner2020operator, bai2021non}. 
However, these methods pose challenges when applied to multi-scale dynamical systems due to the inherent linear characteristics of the model. Such limitation becomes particularly pronounced in the study of plasma dynamics, where transitions can dramatically change in an unpredictable manner due to the complex, nonlinear nature. Therefore, the development of new models is critical to effectively capture the spatiotemporal structures in a low-dimensional space that accurately represents the physical observations. 

Recent advances in scientific machine learning (SciML) present new possibilities to address these challenges in fusion science. Kates-Harbeck et al. proposed the use of recurrent neural networks in FRNN~\cite{kates2019predicting} to forecast disruptions for large burning plasma systems ITER. Recent study on deep reinforcement learning~\cite{degrave2022magnetic} explores an autonomous, self-learning control strategy and produces a set of plasma configurations to control tokamak plasma by acquiring knowledge interacting with its environment. 
The experimental investigations have revealed that the disruptions are associated with the current-driven MHD modes and various additional effects~\cite{Tang2023FRNN}. However, a comprehensive physical understanding of the disruption process remains elusive, and addressing this necessitates the utilization of a simulator capable of elucidating the physics mechanism underlying plasma behaviors at specific time points based on real-time measurement signals. 
The Surrogate of GTC (SGTC) uses deep learning methods to predict plasma instability~\cite{dong2021deep}.
Trained using GTC data, SGTC can accurately predict linear stability and the 2D structure of the kink mode in a few milliseconds --- five orders of magnitude faster than conventional first-principle based simulations.
Nevertheless, as any real fusion plasma system is fundamentally nonlinear, complex and multi-scaled, determining control laws directly from the high-dimensional data is hardly feasible considering its constraints. Therefore, a low-dimensional representation characterizing bifurcation analysis is needed to provide insights into how such complex systems behave, in order to understand and may anticipate critical transitions.

In this work, we introduce a data-driven ROM, Fusion Transfer Learning (FTL) model for real-time characterization and reconstruction of MHD modes. The encoder-decoder based network, as illustrated in Figure~\ref{fig01}, captures the essential spatial features from the GTC simulations of the internal kink modes and is capable of real-time detection of anomalous mode structures when presented with snapshots of dynamical plasma behavior. The network, trained on non-dynamical modes, is extended to embed time-evolution kink instabilities through transfer learning. One of the principal challenges for ROMs is modeling transient system behaviors particularly when the system is activated by smoothly varying external forcing. Trajectories in the latent space provide informative changes in a low-dimensional embedding manner. We focus on identifying the bifurcation point in the low dimensional space observed when plasma instabilities transit from linear to nonlinear phase. 

The remainder of the paper is organized as follows. 
In Section~\ref{sec:results}, we elaborate the results of mode reconstruction, detected anomalous modes, as well as learned nonlinear modes and their elicited bifurcation using the FTL model. 
Section~\ref{sec:discussion} discusses the physical interpretation of the tipping points observed in the latent space, the limitation of this work and potential future directions.
In Section~\ref{sec:methods}, we describe GTC simulation configurations, the architecture of the FTL network, the algorithm to detect anomalous modes, and our approach to learn dynamics through pre-trained models on non-sequential data from linear kink simulations. 

The contributions for this work is summarized as follows: (i) the introduction of FTL, a data-driven ROM, wherein low-dimensional embeddings enable efficient reconstruction of spatiotemoral MHD modes, (ii) development of  method for anomaly filtering and detection based on encoder-decoder network architecture, aiming to identify outliers from raw datasets and improve the fidelity of the pre-trained model, (iii) the demonstration of the proposed FTL on out-of-sample regimes in the parameter space, highlighting its capability for extrapolation and generalization to complex mode structures via transfer learning, (iv) the examination of plasma evolution and dynamic transitions in the latent space, providing insights into the correlation between the encoded representations of dynamic transitions and their manifestations in both real and frequency domains, thereby augmenting the interpretability of the FTL model in physics.
\begin{figure}[!ht]
\begin{center}
\includegraphics[width=0.5\textwidth]{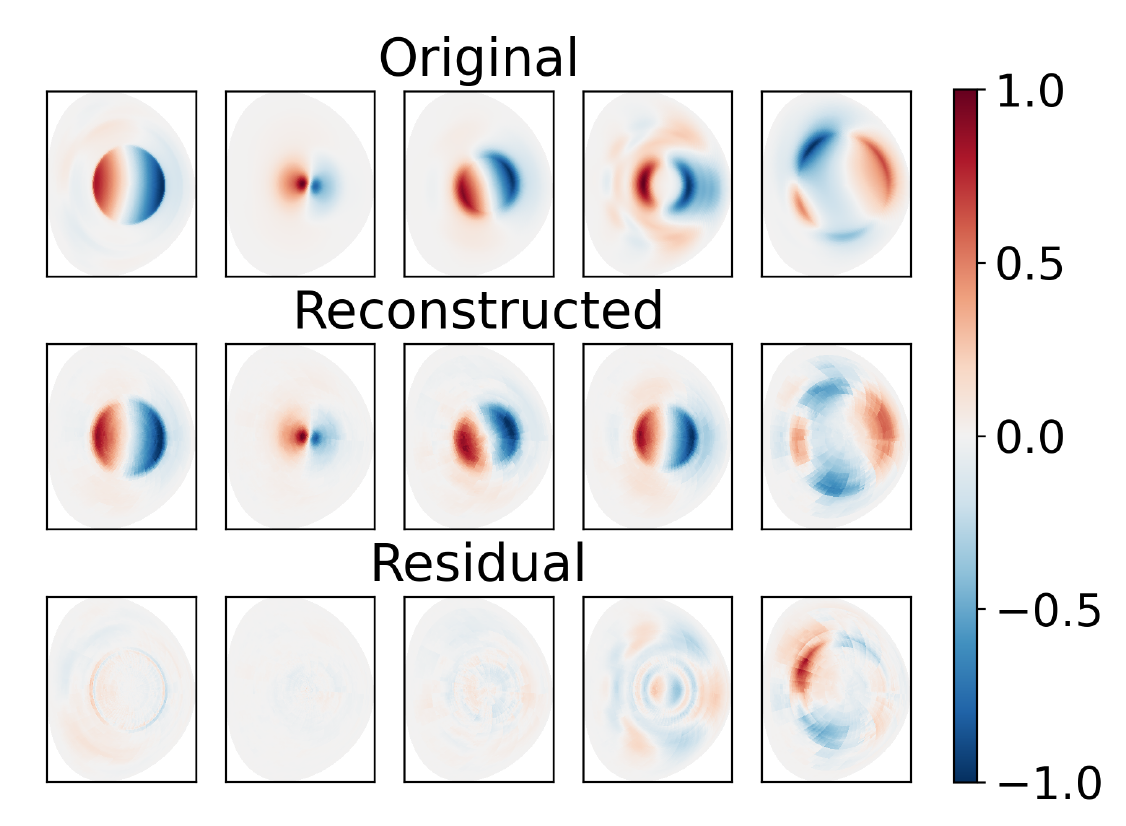}
\caption{Comparison between the original and reconstructed kink modes for five samples from the test data. A larger reconstruction residual is observed for atypical mode structures.} \label{fig02}
\end{center}
\vspace{-.078in}
\end{figure}

\section{Results}\label{sec:results}
We consider a high-dimensional spatio-temporal system $\mathbf{V}(\mathbf{x},t;\mu)$ governed by first-principle PDEs parameterized by $\mu$, and $\mathbf{x}$ and $t$ are spatial and temporal coordinates.
Our goal is to develop a data-driven, transferable ROM that extracts plasma spatial correlations while preserving the dynamics of system in the latent space for a generalized configurations of mode structures.
The simulation data used are  based on the reconstructed equilibrium of DIII-D experimental discharges.
\subsection{Reconstruction and anomaly detection}\label{sec2.1}
In this experiment, we set the free parameter $d=3$ for the dimension of the latent space, and focus on studying the perturbed electrostatic potential in the kink modes. 
Generally, the structure of the kink mode exhibits variability contingent upon a number of factors including magnetic field strength, field line tilt (referred to as the `$q$-profile' as a function of radial position), and the pressure profile. 
The ideal MHD internal kink mode is a current-driven instability with $m=n=1$, where $m$ and $n$ are the poloidal and toroidal mode number. In toroidal devices (e.g., tokamaks), the internal kink mode couples with higher-order poloidal harmonics, thereby altering the stability. The mode structure peaks at the rational surface, characterized by the safety factor $q=1$ in tokamaks. 
The universal feature of the kink mode is the dominant $m=1$ component in the region of $q<1$, accompanied by a subdominant $m=2$ component in the $1<q<2$ region. The $m=0$ component remains constant when approaching to the magnetic axis. We focus our study on the dominant $m = (0,1,2)$ components in this paper while neglecting higher-order components for physical analysis.

\begin{figure*}[!ht]
\begin{center}
\includegraphics[width=.99\textwidth]{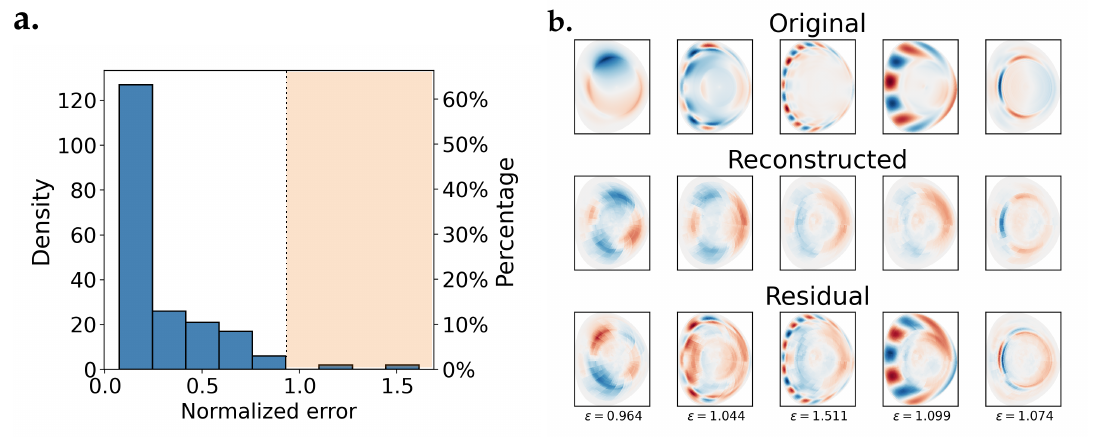}
\caption{Detected anomalous modes using trained FTL network on the test set: (a) a histogram of sample density vs. normalized error  highlighting anomalies in the tail of the distribution, (b) the original, reconstruction, and residual for five kink mode samples in the highlighted in the anomalous (shaded) region of (a), threshold parameter $\tau= 0.95$. }\label{fig03}
\end{center}
\end{figure*}

The specifics of the GTC simulation setup are detailed in Section~\ref{sec:method-GTC}. In all the simulations, the initial conditions are set to $\delta n_e(t=0)=n_\text{init}(r)\cos(\theta-\zeta)$ with $n_\text{init}(r)$ the radial envelope. The envelope peaks at $r=a/2$, where $a$ denotes the minor radius at the plasma wall. And all other quantities are set to $0$ initially. We train our FTL model using $1605$ samples of simulated kink modes obtained from GTC simulations, and test it on another set of $201$ snapshots from unseen data. Note that there is no temporal correlation between the kink mode samples studied in this section, and we investigate dynamical behaviors in Section~\ref{sec:2.2}.
Figure~\ref{fig02} shows five examples of the reconstructed kink modes and their corresponding residuals relative to the original data. General kink mode structures including mode radial location and the different phase angles are well captured via the decoder. The reconstruction error is bigger for atypical structures, such as a dominant $m=2$ structure or a dominant structure near the edge. Unstable kink modes are observed in the first four simulations. The fourth snapshot shows a less common scenario as the `reverse-shear' equilibrium. With increasing radius, the $q$-profile first decreases to a minimum value ($q_{min}$), and then increases. If $q_{min}<1$, there exist two radial positions at which $q=1$, and the $m=1$ structure is mainly located between the two $q=1$ points. The FTL model discovers the structure at the region outside the $q=q_{min}$ flux surface and disregards the structures near the magnetic axis. The fifth instance is a stable case where typical kink structures are absent, resulting in a larger reconstruction error as anticipated. 

As the kink modes are produced from GTC simulation, certain output samples are subjected to subsequent reevaluation and filtration processes. By undergoing extensive training using a substantial dataset comprising numerous kink modes obtained from simulation, the FTL model demonstrates proficiency in reconstructing typical mode structures. This capability enables it to function as an effective filtering mechanism for identifying anomalous modes by examining the deviations when comparing the decoded outputs and original modes, as elucidated in Algorithm~\ref{alg1}. To construct the training dataset, we have cleaned the data via expert labeling and automatic scripts using empirical criteria to exclude the stable or numerical diverging cases.
Figure~\ref{fig03}a presents a histogram illustrating the distribution of normalized reconstruction errors. More than $60\%$ of the test samples resemble the input data with a relative error smaller than $0.2$. The few samples collected in the shaded region that have relatively larger reconstruction errors are shown in Figure~\ref{fig03}b. One of the prevalent anomalies observed is the edge effect, characterized by the aggregation of small-scale structures at the boundary of the poloidal plane. We find these cases are not the physical modes yet failed to be identified by the previous empirical methods. This outcome demonstrates FTL's capacity to discern pollution within the collected data, thereby enhancing the performance of other models trained on the same data. 
Similarly, the trained model can be utilized to identify anomalous modes in real-world experiments, where unexpected structures manifest as outliers in the distribution, detected through the algorithm. 

\subsection{Learning nonlinear dynamics}\label{sec:2.2}
We extend the study of kink modes derived from the linear GTC simulation in Section~\ref{sec2.1} to a nonlinear GTC simulation with kinetic effects. Due to the inclusion of kinetic ion effect, the mode structures rotation with a finite frequency can be observed after the eigenmode is established in the linear phase. This leads to a trajectory rather than a fixed point in the latent space. We observe an abrupt tweak in the modes when the system evolves to the nonlinear phase, and the structure rotates at different speed at different radial positions (i.e., flow shear) because the zonal flow is generated. The flow shear caused by the zonal flow is the main reason for the nonlinear transition and saturation of the mode amplitude. Due to the presence of flow shear, the mode structures are tilted, accompanied by variations in the relative amplitude of different $m$ components when compared to the linear stage.
\begin{figure*}[!t]
\begin{center}
\includegraphics[width=1\textwidth]{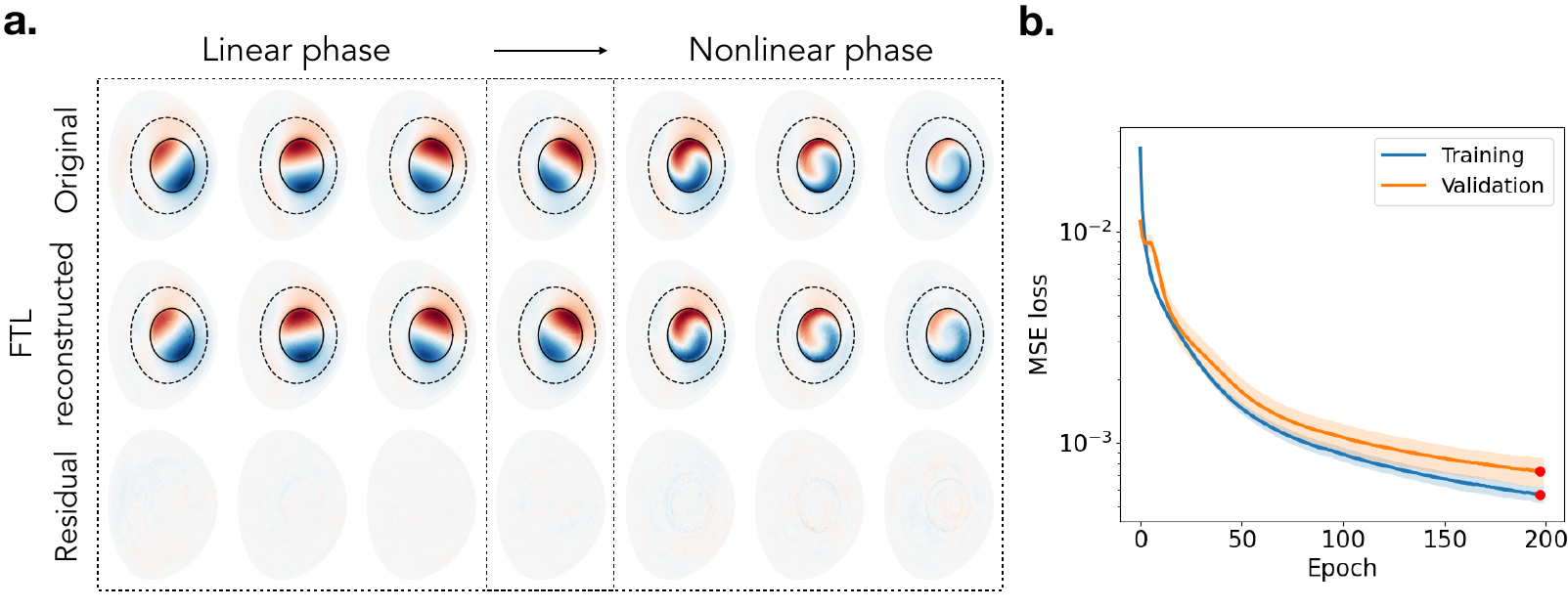}
\caption{(a) Simulated Kink modes with kinetic effects and reconstructed nonlinear mode structures at $t=0.073, 0.135, 0.186, 0.223, 0.250, 0.272, 0.295$ ms using the trained FTL model. (b) Training and validation error vs. epoch number. The solid lines are results averaged over using $20$ different random realizations. The shaded region indicates the distribution of the variance of the reconstruction error in the multiple runs. Early stopping (red dots) is used during the NN training to prevent overfitting.}\label{fig04}
\end{center}
\end{figure*}

\begin{figure*}[!t]
\begin{center}
\includegraphics{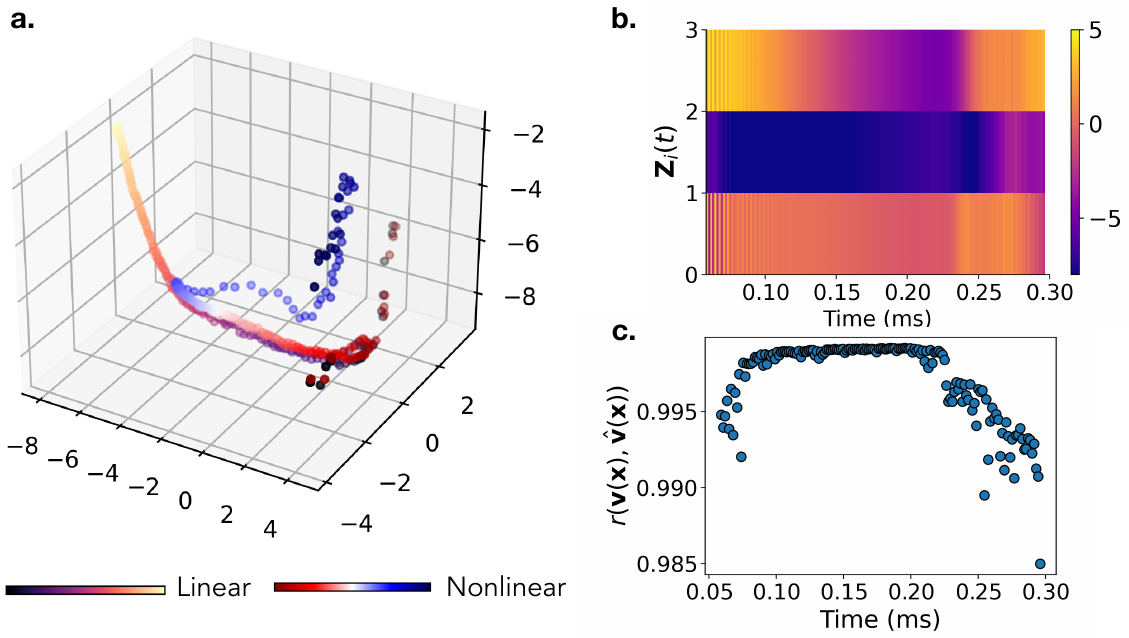}
\caption{(a) Dynamics trajectory of the plasma state evolution in the latent space with time mapped as color. A bifurcation point is observed when the modes transit from linear to nonlinear. (b) latent space vector vs. time. (c) Correlation coefficient of reconstructed modes and ground truth over the time window, $\Delta t_s = 0.5 R_0/C_s = 1.483\mu s$.}\label{fig05}
\end{center}
\end{figure*}

Leveraging the network trained on linear modes from Section~\ref{sec2.1}, we illustrate that the FTL model is able to embed the modes within the latent space and reconstruct the modes accommodating the rotational behaviors in the linear phase. The detailed implementation of the transfer learning process is provided in Algorithm~\ref{alg2}. The nonlinear evolution of the mode structures exhibits distinctive features that are exclusively captured through transfer learning methods. We fine tune the decoder via retraining the FTL network on $128$ snapshots randomly sampled on the data from the nonlinear simulation. Figure~\ref{fig04}a shows the reconstructed modes from the test set, which are temporally correlated to the training set while unseen by the retrained network. The decoded mode structures present strong consistency with the original data, displaying only minor discrepancies. 
As seen in Figure~\ref{fig04}b, the mean-square-error (MSE) loss shows rapid convergence during the network training process for both the training set and test sets, reaching values below $1e-3$ following $100$ epochs. Early stopping is employed to prevent overfitting for generalized performance. The fine-tuned FTL model adeptly reproduces the mode rotation in the linear stage, captures the structure change at the transition point, and accurately depicts the linear structure evolution.

\subsection{Bifurcation analysis}\label{sec:2.3}
As the kink instabilities progress from the linear to nonlinear phase, significant alterations occur in both the characteristics of the mode structures and the rotational state. We investigate the temporal evolution of the kink modes within the low-dimensional embeddings, $\mathbf{Z}(t) = \boldsymbol{\Phi}(\mathbf{V(\mathbf{x},t;\mu)})$ where the spatial domain is compressed to $\mathbb{R}^d$ for each snapshot.

Specifically, we study two cases including one simulation derived from the linear GTC simulation with kinetic effects and the other from the GTC simulation with kinetic and nonlinear effects. The linear and nonlinear simulations start at identical initial conditions. Our investigation focuses on the bifurcation point in the temporal evolution of the kink modes, with the exclusion of the first snapshots under initial effects.
Although the mapping between the latent space coordinates and physical space structures is nonlinear through the encoding layers, which results in the complexity in attributing the effect of a single coordinate, it remains interesting to visualize and examine the relationship between the trajectories in the latent space and their real-space counterparts.
Figure~\ref{fig05}a shows the two dynamic trajectories of the kink mode evolution in the three-dimensional embedded space, where the bifurcation occurs between the linear and nonlinear mode evolution. In the linear stage of the nonlinear simulation, the effect of the nonlinear terms is small, resulting in a system that closely resembles the linear simulation. This can be confirmed by the overlap of the first half of the linear and nonlinear trajectories. 

In the early stage of the simulation, different eigenmodes co-exist and the trajectory shows how the dominant kink eigenmode structure prevails. The compact oscillation between $t=30 R_0/C_s$ and $t=35 R_0/C_s$ may correspond to a physical mode characterized by high frequency and low growth rate. Here, $R_0/C_s$ is used as the time unit, where the major radius at the magnetic axis is $R_0=1.78\text{m}$, and the ion sound speed is $C_s=6.01\times10^5\text{m/s}$. The subsequent smooth evolution in both the trajectories corresponds to the rotation of the most unstable eigenmode in the physical space. When the bifurcation occurs at $t=76.5 R_0/C_s$, the linear trajectory continues to evolve on the consistent path while the nonlinear case deviates from the track sharply and starts to exhibit disturbed behaviors in the latent space. Simultaneously, we note a transition in the mode rotation direction in the physical space of the nonlinear simulation, shifting from clockwise to counterclockwise at the tipping point. The next change in the direction of the nonlinear trajectory may correspond to an increase of the mode tilt. At the later stage, the counterclockwise rotation transits to a minor oscillatory mode where we observe a `reverse' in the trajectory. Overall, examining the dynamics of the latent space proves highly illustrative, revealing distinct physical stages that may not be readily discernible in the real physical space.

The FTL model effectively encodes the coherent mode structures while retaining the intrinsic dynamics within the learned manifold. Figure~\ref{fig05}b illustrates the dynamic changes of the embeddings for $\mathbf{Z}_i$ over the time domain. The first two components, $\mathbf{Z}_1$ and $\mathbf{Z}_2$ represent the positive and negative variations, respectively, of the latent variables; the third component $\mathbf{Z}_3$ captures the transitional points of the instabilities, i.e., when the modes evolve to a stable rotational state at $t=0.10$ ms after the initial perturbations, and the transitional point at bifurcation. The Pearson correlation coefficient $r$ computed between the original snapshot $\mathbf{V}(\mathbf{x})$ and corresponding reconstructed data $\hat{\mathbf{V}}(\mathbf{x})$ across the entire temporal domain is depicted in Figure~\ref{fig05}c, with each snapshot data being centered and scale-invariant,

\begin{align}
r =
  \frac{ \sum_{i=1}^{n}(\mathbf{V}(\mathbf{x})_i-\overbar{\mathbf{V}(\mathbf{x})})(\hat{\mathbf{V}}(\mathbf{x})_i-\overbar{\hat{\mathbf{V}}(\mathbf{x})}) }{%
        \sqrt{\sum_{i=1}^{n}(\mathbf{V}(\mathbf{x})_i-\overbar{\mathbf{V}(\mathbf{x})})^2}\sqrt{\sum_{i=1}^{n}(\hat{\mathbf{V}}(\mathbf{x})_i-\overbar{\hat{\mathbf{V}}(\mathbf{x})})^2}}.
\end{align}
A high linear correlation is observed when eigenmode is formed in the well-defined linear phase, and the coefficient $r$ begins to decrease from $0.998$ to $0.985$ subsequent to the transition point. The linear eigenmode structure is relatively easy to estimate, given the nature of its simple dynamics while we utilize the FTL model to fine tune with temporally correlated data.
The decrease in the correlation aligns with the emergence of mode structures predominantly influenced by the nonlinear effects. Nevertheless, despite this reduction, the reconstruction quality remains acceptable.

\begin{figure*}[ht]
\begin{center}
\vspace{.0925in}
\includegraphics[width=.9\textwidth]{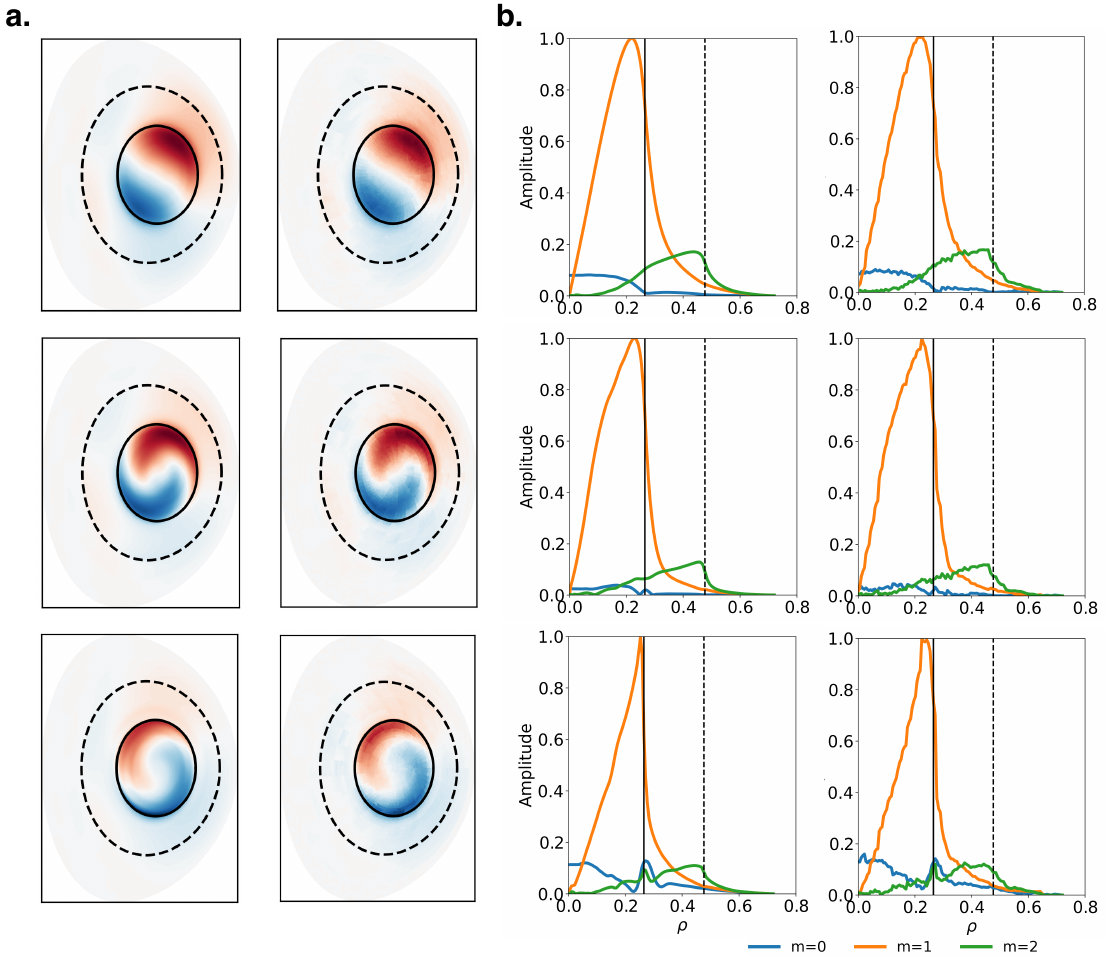}
\caption{2D mode structures and radial structures of the Fourier modes elucidating the transition from linear to nonlinear phase: (a) the original snapshots and reconstructed snapshots at $t = 0.222, 0.237, 0.281$ ms, (b) the amplitude of first three Fourier coefficients $(m=0,1,2)$ along the radius on the poloidal plane at the respective time.}
\label{fig06}
\end{center}
\end{figure*}
\subsection{Towards small or sparse datasets}\label{sec:2.4}
In this section, we evaluate the performance of the FTL model on a limited number of snapshots, which targets real-time tasks for online learning. This is crucial for applications where only a small dataset, characterized by a short temporal span or sparse sampling, is available and expeditious decisions-making is imperative. Extending the FTL architecture, we retrained the network on $17$ snapshots collected from the simulation with a larger sampling interval $\Delta t_s = 5 R_0/C_s$, while maintaining the same initial conditions as in Section~\ref{sec:2.2}.

Figure~\ref{fig06}a shows the original and reconstructed snapshots of the 2D kink mode structures, capturing the transitional behaviors at the bifurcation point and the subsequent nonlinear evolution of modes. The fine-tuned FTL network effectively reconstruct the coherent structures in both the linear and nonlinear phases, exhibiting minor degradation in clarity and sharpness. 
The Fourier decomposition provides a more quantitative comparison; the Fourier coefficients in Figure~\ref{fig06}b show that the FTL reconstruction qualitatively preserves the amplitude and phase of the first three Fourier harmonics. Although the reconstructed structure exhibits minor zigzags, their impact on the physical analysis is negligible. It is remarkable that the reconstructed structures reproduce the critical changes of the Fourier components at the magnetic axis point, i.e., $q=1$ and $q=2$ surfaces precisely, despite that a small amount of loss is observed in the higher order mode.

\begin{figure*}[!ht]
\centering
\includegraphics[width=.99\textwidth]
{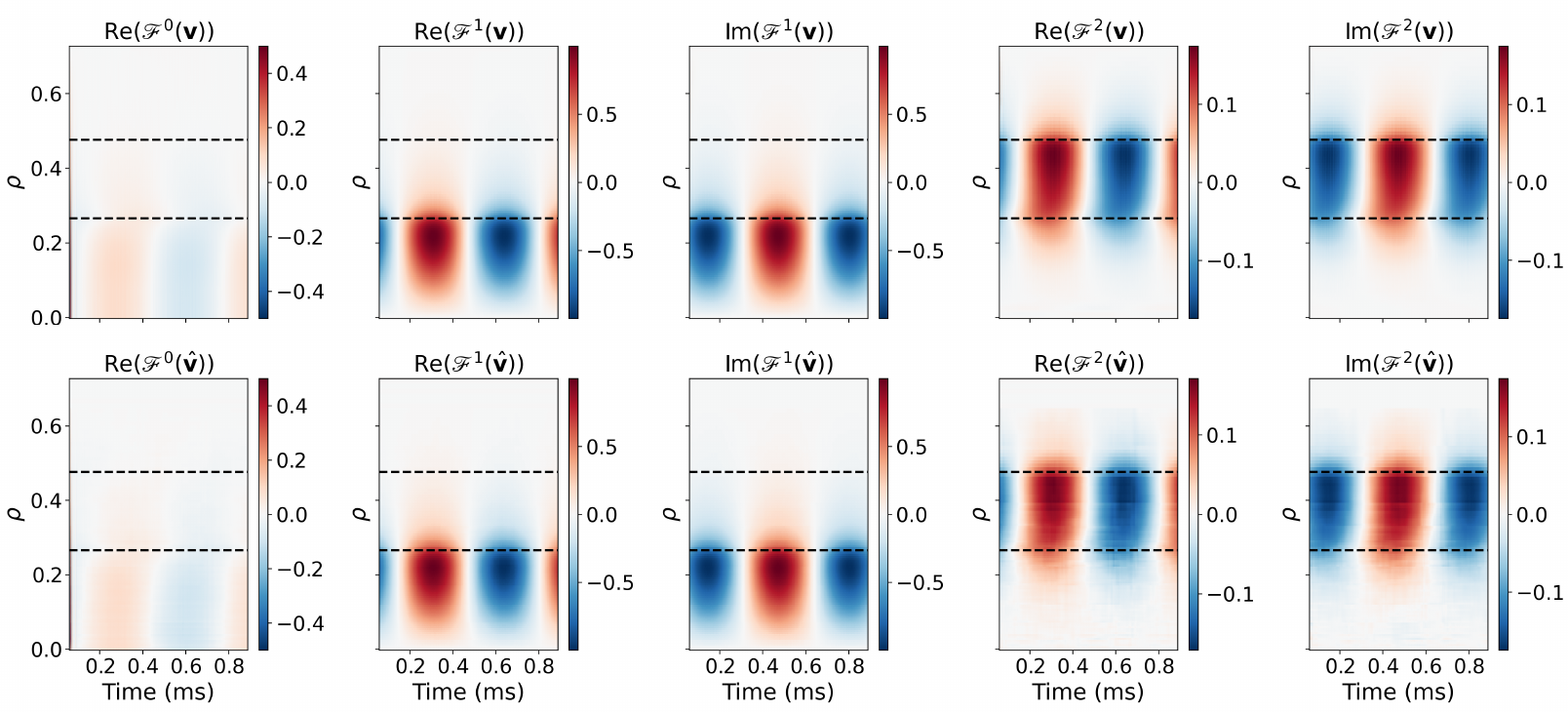}
\caption{Real and Imaginary parts of the Fourier modes for all the snapshots. The reconstructed results (bottom) of the Fourier modes shown good alignment with those obtained from the original GTC data (top). The dash lines mark the $q$ profiles of $q=1,2$ along the radius of the poloidal plane.}
\label{fig07}
\end{figure*}
\begin{figure*}[!ht]
\centering
\includegraphics[width=1\textwidth]{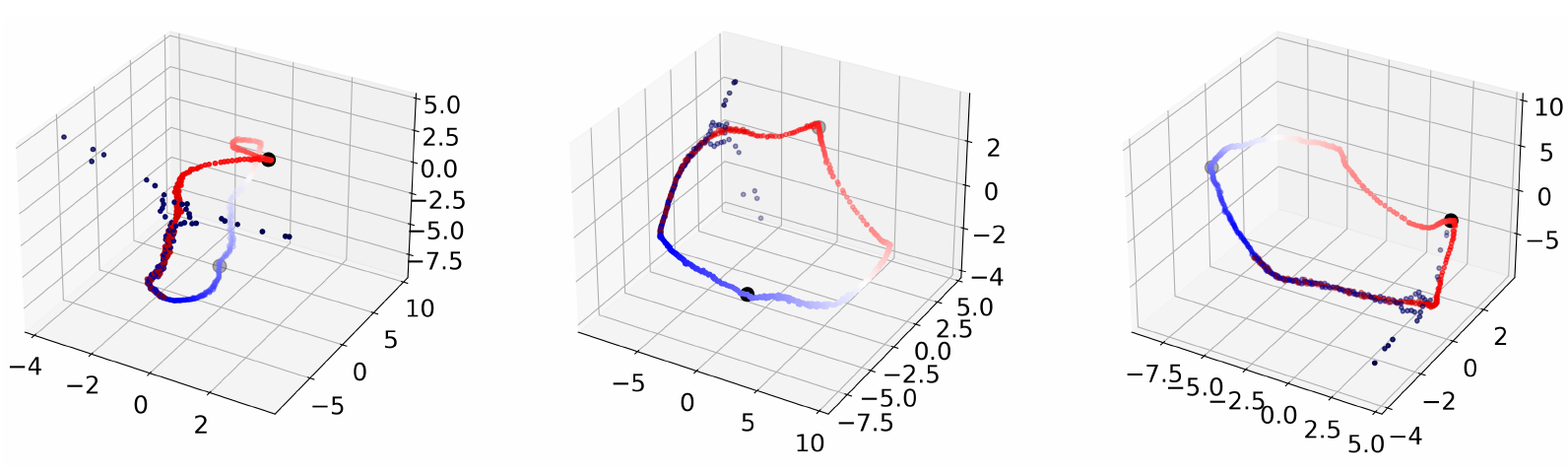}
\caption{Dynamic trajectory of one linear case evolution displayed from three different perspectives in the latent space. At $t=0.665$ ms, a complete evolution period is observed, culminating in the formation of a closed loop. Two tipping points at $t=0.355, 0.651$ ms are highlighted as large black circles, denoting points of notable inflection observed in the evolutionary trajectory.}
\label{fig08}
\end{figure*}
\subsection{Understanding tipping points in latent space}
Interpretation the tipping points in the latent space involves understanding significant transitions or critical thresholds that occur within the underlying data representation. These tipping points may signify abrupt physical state changes, phase transitions, or critical events in the system being studied. Analyzing these points can provide insights into the inherent dynamics, identify regime shifts, or reveal features of importance in the data distribution.

We compare the first three Fourier modes in both their real and imaginary parts in Figure~\ref{fig07}. Periodicity is observed in all the five components: the real parts of 
$\mathscr{F}^0(\mathbf{v}),\mathscr{F}^1(\mathbf{v}),\mathscr{F}^2(\mathbf{v})$, and the imaginary parts of $\mathscr{F}^1(\mathbf{v}), \mathscr{F}^2(\mathbf{v})$.
Figure~\ref{fig08} illustrates the latent space trajectory associated with a linear simulation spanning an extended time evolution. The mode structure in the physical space exhibits a rotational period, which in the latent space corresponds to a closed loop. In traditional methods, an eigenmode can be represented by separating the oscillation over time, the toroidal harmonics, and the poloidal harmonics by carrying out a Fourier transformation. The FTL embeddings provide a low-dimensional representation of the eigenmode, i.e., a visualization of the linear loop trajectory in the encoded latent space. The trajectory in Figure~\ref{fig08} also exhibits dynamical characteristics in addition to the periodic mode rotation. For example, the abrupt direction changes observed at the two flagged points relate to the amplitude of the standing wave component in the physical space. 

\begin{figure}[!t]
\centering
\includegraphics[width=.49\textwidth]{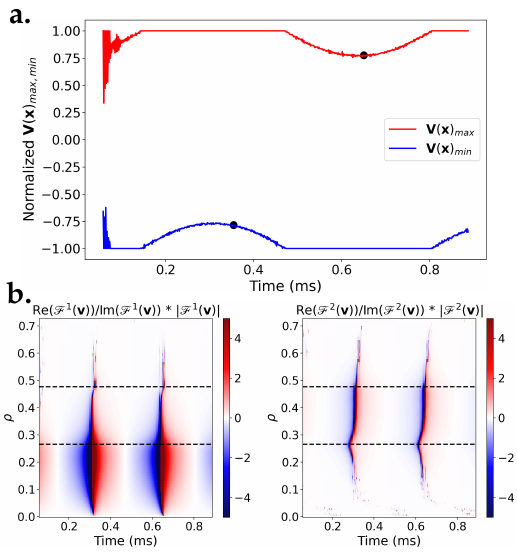}
\caption{The correspondence of the identified tipping points (latent space) in physical space and Fourier domain: (a) normalized spatial maxima and minima for each snapshot in the time domain. (b) the ratio of the real and imaginary parts weighted by the amplitude of the Fourier modes. A phase shift is detected at the same time regime of the transition points as in Fig.~\ref{fig08}.}
\label{fig09}
\end{figure}

The identified tipping points in the latent space are depicted in both physical space and the Fourier domain, demonstrating their correspondence.
Figure~\ref{fig09}a depicts the tipping points observed in the latent space are close to normalized spatial maxima and minima for all the snapshots in the extended linear simulation. The phase shift occurs at approximately $t=0.35$ ms and $t=0.65$ ms in the Fourier domain, as in Figure~\ref{fig09}b, and signifies the transition point where the real-space structure exhibits a distinct change. Specifically, this shift indicates the moment when the pure cosinusoidal wave $\cos(m\theta)$ becomes predominant, while the sinusoidal pattern diminishes. The potential can be generally expressed as 
\begin{multline}
\mathbf{v}=\boldsymbol{\phi}(r,\theta,t)=\sum_{m=0}[(\hat{\boldsymbol{\phi}}_{R,m}(r,t) + i\hat{\boldsymbol{\phi}}_{I,m}(r,t))e^{im\theta} \\+ (\hat{\boldsymbol{\phi}}_{R,m}(r,t) - i\hat{\boldsymbol{\phi}}_{I,m}(r,t))e^{-im\theta}], 
\end{multline}
where $\boldsymbol{\phi}(r,\theta,t)$ is a real number, and the subscripts $R$ and $I$ indicate the real and imaginary parts. 
For a certain mode number $m$, when the imaginary Fourier component is $0$ at $t=t_0$, we have $\boldsymbol{\phi}(r,\theta,t_0)=2\hat{\boldsymbol{\phi}}_{R,m}\cos(m\theta)$. Note that in general $\hat{\boldsymbol{\phi}}_{I,m}(r,t)$ may not be $0$ for all $r$ and all $m$ numbers. In this particular case, the imaginary parts vanish almost altogether near $t=0.35$ ms and $t=0.65$ ms, thus creating the two special tipping points.

%
\section{Discussion}\label{sec:discussion}
We have developed FTL, an encoder-decoder based transfer learning model to represent high-dimensional snapshots of MHD modes in low-dimensional embeddings. Our approach has demonstrated its effectiveness in mode reconstruction, anomaly detection, and revealing transitional dynamical features for unstable kink modes. 
The transferable architecture of FTL, capable of extracting essential information from spatially and/or temporally relevant problems, even with limited datasets, enhances the model's adaptability and efficiency to a spectrum of MHD problems.
The analysis of plasma dynamics in the latent space enables correlations of dynamic transitions with real
and frequency domain behaviors, and adds a layer of physical interpretability to the FTL model.

\textbf{Limitations and future directions.} 
Despite the substantial advances enabled by the FTL model, there are several aspects to be improved upon.  
Whereas this work primarily focuses on mode structures, with snapshot amplitudes being normalized to enhance anomaly detection capabilities, future work will endeavor to incorporate amplitude as well as their growth rates into the FTL model.  Moreover, we envisage leveraging the versatile FTL architecture to encompass multiple physical modes including Neoclassical tearing modes (NTMs) thereby expanding the model's capability. 
One limitation lies in generality of the current hyperparameter settings of FTL's architecture model. Each physical problem necessitates its own hyperpameter tuning, and exploration of network settings, including a network size, depth, activation functions, learning rate, optimizers, batch size, regularization methods and loss functions.

Another avenue for future direction involves addressing concerns regarding the applicability of the current Framework Transfer Learning (FTL) model to unstructured data or irregular geometries, as it is currently designed for structured data with regular geometry. We note that FTL can be extended by replacing the convolutional kernels with graph convolutions in graph neural network (GNNs) in order to process irregular data represented as graphs.  
For specific physics-constrained problems, the loss functions can be formulated with the residual component with respect to the first-principled PDEs, as demonstrated in the physics-informed neural networks~\cite{raissi2019physics, cai2021physics}.  
Additionally, we envision a hybrid approach that integrates FTL with other surrogate models, e.g., SGTC, to develop data-driven embedding methodologies for inferring plasma dynamics. The collaborative framework aims to accurately predict mode structures and their growth rate from constrained encoding, facilitating a deeper understanding and more precise forecasting of disruptions in large-scale plasma systems. 

Ultimately, anticipated future work on model predictive control involves its application in real tokamak experiments, wherein the ongoing study of bifurcation prediction in complex systems necessitates comprehending how system behavior evolves with varying parameters. As bifurcations mark significant shifts in system dynamics, accurate prediction entails a thorough exploration of parameter space to pinpoint transitional regimes. This endeavor will reveal critical conditions triggering bifurcations, providing a comprehensive understanding into the complex systems and guiding strategies for control or mitigation.

\section{Methods}\label{sec:methods}
Consider a full-order model of a nonlinear dynamical system characterized by a system of ordinary differential equations (ODEs), 

\begin{align}
\dot{\mathbf{V}}&=\boldsymbol{f}(\mathbf{V}(\mathbf{x},t;\mu)),\\
\mathbf{V}(\mathbf{x},0;\mu)&=\mathbf{V}^0(\mathbf{x},\mu),
\end{align}
where $t$ denotes time in $\mathbb R_+$, $\mathbf{x}$ denotes the coordinate, $\mathbf{V}$ denotes the state variable, $\mu \in \mathcal{D}$ denotes the system parameters, and $\mathbf{x}^0(\mu)$ is the parameterized initial condition. $\boldsymbol{f}: \mathbb{R}^N \times \mathbb R_+ \times \mathcal{D} \rightarrow \mathbb{R}^N$ denotes the dynamics of the system.
\subsection{Encoder-decoder network}
The proposed FTL consists of convolutional, fractionally-strided deconvolutional blocks and fully-connected layers built in an autoencoder architecture. The convolutional kernels employed in the FTL network are adept at capturing spatial correlations and extracting hierarchical patterns from training data. The encoder layers constrains coherent spatial patterns in the bottleneck layers. The operator ${\Phi}:\mathbb{R}^N \rightarrow \mathbb{R}^d$ maps the state $\mathbf{V}$ to a low-dimensional embeddings $\mathbf{Z}$ through convolutional kernels and a feed-forward neural network. The constructed latent space encodes the spatial structures while preserving the dynamical characteristics of the system when trained on a series of snapshots. The dimension of the latent space $d$ is a hyperparameter that can be evaluated based on the complexity of the system and trade-off between compression ratio and reconstruction quality. The decoder ${\Psi}$ network reconstruct the original input from the encodings through fully-connected layers and deconvolutional kernels, ${\Psi}: \mathbb{R}^d \rightarrow \mathbb{R}^N$. Bath normalization layers are used as a regularizer to improve the training stability and prevent overfitting. We choose the Rectified Linear Unit (Relu) activation function mapping the nonlinearities among the hidden layers, and the hyperbolic tangent function to constrain the output variables to the range of $[-1,1]$. We compute the parameters of the network by minimizing the Frobenius norm of the loss function over all the batches during the stochastic gradient descent (SGD) of (re)training, 

\begin{align}\label{eq4}
{l} = \mathcal{\mathbb{E}}(\lVert \mathbf{V}(\mathbf{x}, t;\mu)-\Psi\circ\Phi(\mathbf{V}(\mathbf{x}, t;\mu))\rVert)_F^2 + \lambda\lVert\omega \rVert_2^2.
\end{align}
As the model learns to reconstruct the input data while minimizing the reconstruction error, we set up a threshold $\tau$ based on the percentile of the distribution of the relative residual

\begin{align}
\epsilon_r(\mathbf{v}) = \frac{\lVert \mathbf{v}(\mathbf{x};\mu)-\Psi\circ\Phi(\mathbf{v}(\mathbf{x};\mu))\rVert)_F} {\lVert  \mathbf{v}(\mathbf{x};\mu)\rVert_F},
\end{align}
and separate normal samples from anomalies. 
Algorithm~\ref{alg1} provides a summary of the FTL offline training, including the anomaly filtering procedure. Network hypeparameters, including the learning rate $\eta$, batch size $n_\text{batch}$, maximum number of epochs $n_\text{epoch}$ are tuned to optimize the performance on the validation sets. The regularization term $\lambda\lVert\omega \rVert_2^2$ penalizes large weights in the model to prevent overfitting for improving generalization performance. Adjusting the parameter $\lambda$ controls the strengths of the regularization applied to the model.
Early stopping is also recommended as a measure to mitigate potential overfitting. The level of anomaly filtering is determined by the threshold parameter $\tau$. It is important to note that the filtering process can be iterative in hierarchical anomaly detection, especially when multiple levels of granularity or complexity exist within the system. This iterative approach allows for a comprehensive examination of anomalies across different levels of abstraction, ensuring a thorough identification process. Following the removal of anomalous samples, we proceed to retrain the FTL network using the normal samples. This retraining step aims to improve the model quality in preparation for subsequent tasks. 
The trained model possesses transferable capabilities that enables the acquisition low-dimensional representations from spatially and/or temporally relevant systems when provided with a few new samples. 
This function notably benefits problems with limited data availability or in the study of intricate systems where collecting a substantial number of samples is infeasible.

\begin{algorithm}
\caption{FTL offline training and anomaly filtering}\label{alg1}
\begin{algorithmic}[1]
\INPUT Network architecture; training set $\mathcal{T}$; validation set $\mathcal{V}$; anomaly threshold $\tau$; NN hyperparameters (learning rate $\eta$; batch size $n_\text{batch}$; maximum number of epochs $n_\text{epoch}$; regularization parameter $\lambda$; early-stopping criterion $k$).
\State Standardize data in the training and validation set;
\State Initialize iterations $i \gets 0$ and initial parameters $\omega^0;$
\While{$i \le n_\text{epoch}$}
    \For{$j=1,\cdots,n_\text{batch}$}
        \State $\omega^{i+1} \gets \omega^i$; \Comment{update weights}
    \If{$l_{val}^i > \forall_{i\ge k}{l_{val}^{i-k}}$} 
        \State $\omega^* \gets \omega^{i-k};$ \Comment{early stopping}
        \State \textbf{break};
    \EndIf
    \EndFor
\EndWhile
\State Set ${\Phi} \gets {\Phi}(\omega^*)$, ${\Psi} \gets {\Psi}(\omega^*)$; \Comment{save operators}
\For{$\mathbf{v} \in \mathcal{T}$}
    \If{$\epsilon_r(\mathbf{v}) > \tau$}
        \State $\mathcal{T} \gets \mathcal{T} \backslash \mathbf{v};$ \Comment{filter out anomalies}
    \EndIf
\EndFor
\OUTPUT Encoder ${\Phi}$; decoder ${\Psi}$; refined training set $\mathcal{T}$.
\end{algorithmic}
\end{algorithm}

Algorithm~\ref{alg2} outlines the process of leveraging the pretrained FTL model to adapt to a new dataset through transfer learning. 
After standardizing the data, the initial parameters obtained from the trained model in Algorithm~\ref{alg1} are utilized as the starting configuration. Subsequently, the network weights (and biases) $\omega$ undergo updates in each iteration to fit the sample data from the new problem. Once the convergence of the loss function for reconstruction reaches a satisfactory level, anomaly filtering can be similarly applied.
This approach is particularly advantages for handling snapshots characterized by spatial or temporal correlations in online tasks. By employing the fine-tuning method, the computational complexity and training time are significantly reduced, allowing for efficient model adaptation. More importantly, the utilization of an existing trained model enables rapid convergence, even with a minimal number of samples, which would otherwise present challenges if starting the training process from scratch. 
The architecture of FTL utilized in this work is illustrated in the Figure~\ref{figs01}.

\begin{algorithm}
\caption{FTL online fine-tuning}\label{alg2}
\begin{algorithmic}[1]
\INPUT Network architecture; online set $\mathcal{S}$; encoder $\Phi$; decoder $\Psi$; NN hyperparameters (learning rate $\eta$; batch size $n_\text{batch}$; maximum number of epochs $n_\text{epoch}$; regularization parameter $\lambda$; early-stopping criterion $k$); (optional) anomaly threshold $\tau$.
\State Randomly split the online set $\mathcal{S}$ into training $\mathcal{S}_{tr}$ and validation set $\mathcal{S}_{val}$;
\State Standardize data in the training and validation set;
\State Initialize iterations $i \gets 0;$
\State Set encoder parameters $\omega_e=\omega(\Phi);$
\State Initialize decoder parameters $\omega_d^0=\omega(\Psi);$
\While{$i \le n_\text{epoch}$}
    \For{$j=1,\cdots,n_\text{batch}$}
    \If{$l_{val}^i > \forall_{i\ge k}{l_{val}^{i-k}}$} 
        \State $\omega_d^* \gets \omega_d^{i-k};$ 
        \State \textbf{break};
    \EndIf
    \EndFor
\EndWhile
\State Define $\mathbf{Z} \gets \Phi(\mathbf{V}_\mathcal{S}(\mathbf{x}));$ \Comment{encode variables}
\State $\hat{\mathbf{V}}_\mathcal{S}(\mathbf{x})\gets \Psi(\omega_d^*)\circ\Phi(\mathbf{V}_\mathcal{S}(\mathbf{x});$ \Comment{decode embeddings}
\Procedure{Optional }{Anomaly set $\mathcal{A}$} 
\For{$\mathbf{v} \in \mathcal{S}$}
    \If{$\epsilon_r(\mathbf{v}) > \tau$}
        \State $\mathcal{A} \gets \mathcal{A} \cup \mathbf{v};$
    \EndIf
\EndFor
\EndProcedure
\OUTPUT Embeddings $\mathbf{Z};$ reconstructed snapshots $\hat{\mathbf{V}}_\mathcal{S}(\mathbf{x})$; (optional) anomaly set $\mathcal{A}$.
\end{algorithmic}
\end{algorithm}


\subsection{GTC simulation model and simulation settings}\label{sec:method-GTC}
The GTC model used to simulate the $5000$ DIII-D experimental shots~\cite{dong2021deep} is the single-fluid model in low-frequency, long-wavelength limit~\cite{wei2021electromagnetic}. GTC uses the perturbation method, in which the physical quantities are separated to equilibrium and perturbed parts. The equilibrium quantities, including temperature, density, magnetic field and the flux surface shape are taken from the reconstruction of DIII-D experiments, and the perturbed quantities are solved from GTC equations. The first equation is the continuity equation for gyrocenter charge density,

\begin{align}
\begin{split}
 & \frac{\partial\delta n}{\partial t}+\mathbf{B}_{0}\cdot\nabla\left(\frac{n_{0}\delta u_{\parallel}}{B_{0}}\right)-n_{0}\mathbf{v}_{*}\cdot\frac{\nabla B_{0}}{B_{0}}+\delta\mathbf{B}_{\perp}\cdot\nabla\left(\frac{n_{0}u_{\parallel0}}{B_{0}}\right)\\
 & -\frac{\nabla\times\mathbf{B_{0}}}{eB_{0}^{2}}\cdot\left(\nabla\delta P_{\parallel}+\frac{\left(\delta P_{\perp}-\delta P_{\parallel}\right)\nabla B_{0}}{B_{0}}\right)\\
 & +\nabla\cdot\left(\frac{\delta P_{\parallel}\mathbf{b}_{0}\nabla\times\mathbf{b}_{0}\cdot\mathbf{b}_{0}}{eB_{0}}\right) -\frac{\mathbf{b}_{0}\times\nabla\delta B_{\parallel}}{e}\cdot\nabla\left(\frac{P_{0}}{B_{0}^{2}}\right)\\
 &-\frac{\nabla\times\mathbf{b}_{0}\cdot\nabla\delta B_{\parallel}}{eB_{0}^{2}}P_{0}=0,\label{eq:MHD_continuity}
\end{split}
\end{align}
where $n$ is the density, $B$ is the magnetic field, $u_\parallel$ is the parallel flow velocity, $P$ is the pressure. The subscript $0$ denotes the equilibrium quantity, and the equilibrium is prescribed in a single GTC simulation. The $\delta$ prefix means the perturbed quantity, the evolution of which is calculated through the simulation. $e$ is the elementary charge, $e=-q_e$. $\mathbf{b}_0=\mathbf{B}_0/B_0$ is the unit vector in the direction of magnetic field line. The subscript $\parallel$ and $\perp$ means the component parallel and perpendicular to the magnetic field of a vector. Note here $\delta n=\delta n_e + q_i\delta n_i/q_e$ stands for the difference of ion and electron density, and $\delta u_\parallel=\delta u_{\parallel e} +q_i \delta u_{\parallel i}/q_e$ is the difference of ion and electron flow. $\mathbf{v}_{*} = \mathbf{b}_{0}\times\nabla\left(\delta P_{\parallel}+\delta P_{\perp}\right)/\left(n_{0}m_{e}\Omega_{e}\right)$. $m_e$ is the electron mass, $\Omega_e=eB_0/m_e$ is the electron cyclotron frequency. The perturbed electron parallel flow $\delta u_\parallel$ can be solved from Ampere's law,

\begin{equation}
    \delta u_{\parallel}=\frac{1}{\mu_{0}en_{0}}\nabla_{\perp}^{2}\delta A_{\parallel},\label{eq:Ampere_MHD}
\end{equation}
where $\mu_0$ is the permeability of vacuum. $\delta A_\parallel$ is the perturbed vector potential. In the single fluid model, $E_\parallel = 0$ is assumed. Then $\delta A_\parallel$ can be solved from

\begin{equation}
    \frac{\partial A_{\parallel}}{\partial t}=\mathbf{b}_{0}\cdot\nabla\phi,\label{eq:Apara_MHD}
\end{equation}
and the electrostatic potential $\phi$ can be solved from gyrokinetic Poisson's equation (the quasi-neutrality condition)

\begin{equation}
    \frac{c^{2}}{v_{A}^{2}}\nabla_{\perp}^{2}\phi=\frac{e\delta n}{\epsilon_{0}},\label{eq:poisson_MHD}
\end{equation}
 where $c$ is the speed of light, $v_{A}$ the Alfv\'{e}n velocity, and $\epsilon_{0}$ the dielectric constant of vacuum. The parallel magnetic perturbation $\delta B$ is given by the perpendicular force balance,
 
\begin{equation}
\frac{\delta B_{\parallel}}{B_{0}}=-\frac{\beta_e}{2}\frac{\delta P_\perp}{P_{0}}=-\frac{\beta_{e}}{2}\frac{\partial P_{0}}{\partial\psi_{0}}\frac{\delta\psi}{P_{0}}.\label{eq:bpara_MHD}
\end{equation}
The perturbed pressure in the fluid limit can be calculated by

\begin{align}
\delta P_{\perp} & =\frac{\partial P_{0}}{\partial\psi_{0}}\delta\psi-2\frac{\delta B_{\parallel}}{B_{0}}P_{0},\label{eq:Pperp_ad_MHD}
\end{align}

\begin{align}
\delta P_{\parallel} & =\frac{\partial P_{0}}{\partial\psi_{0}}\delta\psi-\frac{\delta B_{\parallel}}{B_{0}}P_{0}.\label{eq:Ppara_ad_MHD}
\end{align}
In these equations, $\psi_0$ and $\delta\psi$ is the equilibrium and perturbed magnetic flux, and the evolution of $\delta\psi$ is solved from

\begin{align}
    \frac{\partial\delta\psi}{\partial t} & =-\frac{\partial\phi}{\partial\alpha_{0}},\label{eq:dpsi}
\end{align}
where $\alpha_0$ is from the Clebsch representation of $\mathbf{B}$ field, $\mathbf{B}_0=\nabla\psi_0\times\nabla\alpha_0$. For all the 5000 simulations in DIII-D tokamak geometry, we use the $100\times250\times24$ mesh grids in the radial, poloidal and parallel direction. The time step is set to $\Delta t=0.01R_0/C_s$, where $R_0$ is the distance between magnetic axis to the geometric center of the tokamak, $C_s=\sqrt{T_e/m_i}$ is the acoustic velocity. The physical time of $\Delta t$ depends specific parameters. For typical DIII-D parameters, $R_0\approx1.6 \text{m}$, $T_e\approx10 \text{keV}$, so $\Delta t\approx1.6\times10^{-8} \text{s}$. The total number of simulation steps is set to 3000. Due to the free energy contained in the pressure gradient and the parallel current, the internal kink mode can be driven unstable. The volume averaged mode amplitude is measured to calculate the linear mode growth rate, and a number of snapshots during the simulation are taken to study the mode structure. Only one toroidal mode number $n=1$ is kept in the simulations. There are $1972$ cases of the simulated kink modes identified as unstable. We use several empirical methods to peel off the cases with large numerical noise, and the total number used to train the FTL model is $1605$. The cases with stable kink modes cannot show physical kink mode structure, therefore not used in this paper.

Apart from the fluid simulations of the mentioned dataset, we also ran the linear and nonlinear gyrokinetic simulation of DIII-D discharge \#141216 at $t = 1750$ ms for this paper. The ions are described by Vlasov equation,

\begin{equation}
Lf_{s}\equiv\left(\frac{\partial}{\partial t}+\dot{\mathbf{R}}\cdot\nabla+\dot{v}_{\parallel}\frac{\partial}{\partial v_{\parallel}}\right)f_{i}\left(\mathbf{R},\mu,v_{\parallel}\right)=0,
\end{equation}
where $\mathbf{R}$ is the ion gyrocenter position, $\mu$ is the gyrokinetic magnetic moment, $v_\parallel$ is the parallel velocity, $f_i$ is the ion gyrokinetic distribution function. The gyroaveraged ion gyrocenter density, flow, and pressure can be obtained by integrating the distribution function,

\begin{equation}
\delta\bar{n}_{i}\left(\mathbf{x}\right)=\int d\mathbf{v}\int d\mathbf{R}\delta f_{i}\left(\mathbf{R}\right)\delta\left(\mathbf{x}-\mathbf{R}-\boldsymbol{\rho}\right),
\end{equation}
\begin{equation}
n_{i}\delta\bar{u}_{\parallel i}\left(\mathbf{x}\right)=\int d\mathbf{v}v_{\parallel}\int d\mathbf{R}\delta f_{i}\left(\mathbf{R}\right)\delta\left(\mathbf{x}-\mathbf{R}-\boldsymbol{\rho}\right),
\end{equation}

\begin{equation}
\delta P_{\perp i}\left(\mathbf{x}\right)=\int d\mathbf{v}\mu B_0\int d\mathbf{R}\delta f_{i}\left(\mathbf{R}\right)\delta\left(\mathbf{x}-\mathbf{R}-\boldsymbol{\rho}\right),
\end{equation}
where $\boldsymbol{\rho}$ is the gyroradius, $\delta f_i = f_i-f_{0i}$, and $f_{0i}$ are assumed to be the Maxwellian distribution function. The electrons are still described by the fluid equation,

\begin{align}
\begin{split}
	&\frac{\partial\delta n_{e}}{\partial t}+\mathbf{B}_{0}\cdot\nabla\left(\frac{n_{0e}\delta u_{\parallel e}}{B_{0}}\right)+B_{0}\mathbf{v}_{E}\cdot\nabla\left(\frac{n_{0e}}{B_{0}}\right)\\
 &-n_{0}\left(\mathbf{v}_{*}+\mathbf{v}_{E}\right)\cdot\frac{\nabla B_{0}}{B_{0}}+\delta\mathbf{B}_{\perp}\cdot\nabla\left(\frac{n_{0e}u_{\parallel0e}}{B_{0}}\right)\\
	&-\frac{\nabla\times\mathbf{B_{0}}}{eB_{0}^{2}}\cdot\left(\nabla\delta P_{\parallel e}+\frac{\left(\delta P_{\perp e}-\delta P_{\parallel e}\right)\nabla B_{0}}{B_{0}}-n_{0e}e\nabla\delta\phi\right)\\
 &+\nabla\cdot\left(\frac{\delta P_{\parallel e}\mathbf{b}_{0}\nabla\times\mathbf{b}_{0}\cdot\mathbf{b}_{0}}{eB_{0}}\right) +\delta\mathbf{B}_{\perp}\cdot\nabla\left(\frac{n_{0e}\delta u_{\parallel e}}{B_{0}}\right)\\
 &+B_{0}\mathbf{v}_{E}\cdot\nabla\left(\frac{\delta n_{e}}{B_{0}}\right)+\frac{\delta n_{e}}{B_{0}^{2}}\mathbf{b}_{0}\times\nabla B_{0}\cdot\nabla\phi\\
 &+\frac{\delta n_{e}}{B_{0}^{2}}\nabla\times B_{0}\cdot\nabla\phi-\frac{\mathbf{b}_{0}\times\nabla\delta B_{\parallel}}{e}\cdot\nabla\left(\frac{\delta P_{\perp e}+P_{0e}}{B_{0}^{2}}\right)\\
&-\frac{\nabla\times\mathbf{b}_{0}\cdot\nabla\delta B_{\parallel}}{eB_{0}^{2}}\left(\delta P_{\perp e}+P_{0e}\right)=0,\label{eq:Electron_Continuity}
\end{split}
\end{align}
Note this electron continuity equation resembles \eqref{eq:MHD_continuity}, but includes the nonlinear terms and $\mathbf{v}_E$ terms which describes the $\mathbf{E}\times\mathbf{B}$ velocity. We separate the electron parallel flow into zonal ($m=n=0$) and non-zonal part, and the non-zonal part is given through

\begin{equation}
en_{e}\delta u_{\parallel e, nz}=\frac{1}{\mu_{0}}\nabla_{\perp}^{2}\delta A_{\parallel, nz}+Z_{i}n_{i}\delta\bar{u}_{\parallel i, nz}. \label{eq:uepara}
\end{equation}
We define $E_{\parallel,nz}=-\nabla_\parallel\phi-\frac{\partial \delta A_{\parallel,nz}}{\partial t} \equiv-\nabla_\parallel\phi_{eff}$, then $\delta A_{\parallel, nz}$ is calculated from
\begin{equation}
    \frac{\partial A_{\parallel,nz}}{\partial t}=\mathbf{b}_{0}\cdot\nabla(\phi_{eff}-\phi_{nz}),\label{eq:Apara}
\end{equation}
where $\phi_{eff}$ is given by

\begin{equation}
\frac{e\phi_{eff}}{T_{e}}=\frac{\delta n_{e}}{n_{0e}}+\frac{\delta B_{\parallel}}{B_{0}}-\frac{\partial\ln n_{0}}{\partial\psi_{0}}\delta\psi.\label{eq:phieff_ad}
\end{equation}
The electrostatic potential and parallel magnetic perturbation are solved from the coupled equations,

\begin{align}
\begin{split}
&\frac{Z_{i}^{2}n_{i}}{T_{i}}\left(\phi-\tilde{\phi}_{i}\right)\\
&-\frac{1}{B_{0}}\left(Z_{i}n_{i0}\left\{ \delta B_{\parallel}\right\} _{i}-en_{0}\left\{ \delta B_{\parallel}\right\} _{e}\right)\\
=&Z_{i}\bar{n}_{i}-en_{e},\label{eq:Poisson_orig}   
\end{split}
\end{align}
\begin{equation}
\begin{aligned}
    \frac{\delta B_\parallel}{B_0}
    =&-\frac{\beta_e}{2+\beta_e+2\beta_i}\left[\frac{3}{2}\nabla_\perp^2\phi\frac{\beta_i}{\beta_e}\frac{Z_i}{T_i}\rho_i^2\right.\\
    &+\frac{5}{4}\nabla_\perp^4\phi\frac{\beta_i}{\beta_e}\frac{Z_i}{T_i}\rho_i^4\\
    &\left.+\frac{1}{P_{0e}}\left(\delta n_eT_e + n_{0e}\frac{\partial T_e}{\partial\psi}\delta\psi+\delta P_{\perp i}\right)\right], \label{eq:Bpara}
\end{aligned}
\end{equation}
where $\beta_{s}=2\mu_0P_{0s}/{B_0^2}$, and $\delta\tilde{\phi}$ is the double averaged potential,

\begin{equation}
\tilde{\phi}\left(\mathbf{x}\right)=\int d\mathbf{v}\int d\mathbf{R}\left\langle \phi\right\rangle\left(\mathbf{R}\right)f_{0}\left(\mathbf{R}\right)\delta\left(\mathbf{x}-\mathbf{R}-\boldsymbol{\rho}\right),
\end{equation}
and 

\begin{equation}
\begin{split}
    \left\langle \phi \right\rangle\left(\mathbf{R}\right)=\frac{1}{2\pi}\oint d\zeta\int d\mathbf{x}\phi\left(\mathbf{x}\right)\delta\left(\mathbf{x}-\mathbf{R}-\boldsymbol{\rho}\right).
\end{split}
\end{equation}
$\left\{ \delta B_{\parallel}\right\} _{i}$ denotes the double gyro-averaged
$\delta B_{\parallel}$,

\begin{align}
\begin{split}
    \left\{ \delta B_{\parallel}\right\} _{i}\left(\mathbf{x}\right) =& \frac{m\Omega_{i}^{2}}{2\pi n_{0s}T_{i}}\int d\mathbf{v}\int d\mathbf{R}\int_{0}^{\rho}r'dr'\int_{0}^{2\pi}d\zeta'\\
    &\int d\mathbf{x'}\delta B_{\parallel}\left(\mathbf{x}'\right)\delta\left(\mathbf{x}'-\mathbf{R}-\mathbf{r'}\right)\\
 & \times f_{0s}\delta\left(\mathbf{x}-\mathbf{R}-\boldsymbol{\rho}\right),
\end{split}
\end{align}
and $\zeta$ is the gyromotion phase angle. Finally, we have the equation for $\delta\psi$,

\begin{align}
        \frac{\partial\delta\psi}{\partial t} & =-\frac{\partial(\phi_{eff}-\phi_{nz})}{\partial\alpha_{0}},\label{eq:dpsi2}
 \end{align}
 and an additional nonlinear part for $A_\parallel$,
 
 \begin{equation}
\begin{aligned}
    \frac{\partial A_\parallel^{NL}}{\partial t} =& - \frac{m_e}{n_0 e^2}\nabla\cdot\left(\delta u_{\parallel e} \frac{P_{0e}\mathbf{B}_0\times\nabla\delta B_\parallel}{B_0^3}\right) \\
    & + \frac{\delta\mathbf{B}}{B_0}\cdot\nabla(\phi_{eff}-\phi_{nz}) + \frac{P_{0e}}{en_0}\frac{\delta\mathbf{B}_\perp}{B_0^2}\cdot\nabla\delta B_{\parallel}.
    \end{aligned}
\end{equation}
This $\partial A_\parallel^{NL}$ term includes the nonlinear part of $A_\parallel$, and the zonal component of $A_\parallel$ as well. In this paper, we use the same grid number $100\times250\times24$. The time step is set to $\Delta t=0.005R_0/C_s = 1.483\times 10^{-8} \text{s}$, where $R_0/C_s$ represents a commonly used time unit in tokamak simulations, with $R_0$ the major radius at magnetic axis, and $C_s$ the ion sound speed. 
We run $20000$ steps for the nonlinear simulation, in which we can clearly see the nonlinear saturation and evolution. We keep both the $n=1$ and $n=0$ modes in the simulation. In the linear simulation, all nonlinear terms are forced to be $0$, and we run $60000$ steps until we see the clear periodic behaviors. In this paper, we only analyze the dynamics of electrostatic potential mode structure.

\section{Data availability}
All data needed to evaluate the conclusions in the paper are present in the paper. Additional data related to this paper are available from the corresponding author upon reasonable request.
\section{Code availability}
All the source codes that were used for the findings of this study will be openly available in the GitHub at https://github.com/zhbai/FTL upon publication.

\section*{Acknowledgements}
ZB, WT, LO and SW gratefully acknowledge support from the U.S. Department of Energy, Office of Science, SciDAC/Advanced Scientific Computing Research under Award Number DE-AC02-05CH11231. XW, WT and ZL acknowledge support from the U.S. Department of Energy, SciDAC/Fusion Energy Sciences and the INCITE award.
This research used resources of the National Energy Research Scientific Computing Center (NERSC), a U.S. Department of Energy Office of Science User Facility located at Lawrence Berkeley National Laboratory, operated under Contract Number DE-AC02-05CH11231.

\section*{Supplementary Information}
\begin{figure*}[!b]
\begin{center}
\includegraphics[width=1\textwidth]{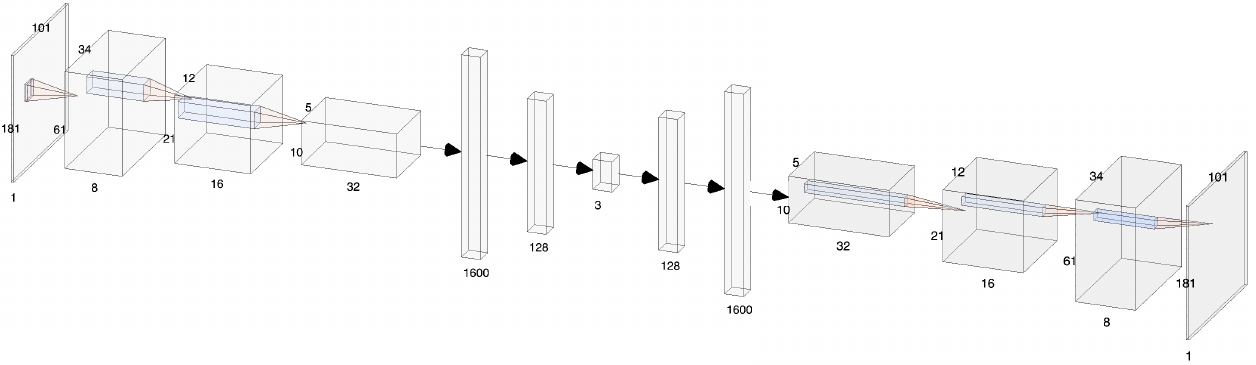}
\end{center}
\caption{Schematics of the FTL network used in this work that consists of convolutional, fractionally-strided deconvolutional blocks and fully-connected layers built in an autoencoder architecture.}\label{figs01}
\end{figure*}

Figure~\ref{figs01} illustrates the architecture of FTL network used in this work.

\bibliographystyle{siam}
\bibliography{Reference}

\end{document}